# Novel two-dimensional materials for electronics: computational modelling as a tool to enable the use of phosphorene in real applications

Francesco Mercuri*

*Consiglio Nazionale delle Ricerche (CNR), Istituto per lo Studio dei Materiali Nanostrutturati (ISMN), Via P. Gobetti 101, 40129 Bologna, Italy*

francesco.mercuri@cnr.it

Two-dimensional and layered materials, such as graphene, have emerged in recent years for their potential use in several applications in technology, for example in electronics, bioelectronics, optoelectronics and related fields. Phosphorene, which can be considered as an analogous of graphene made of phosphorus atoms, is a relatively new two-dimensional system with peculiar structural, electronic, and mechanical properties. Despite the remarkable potential for applications, phosphorene suffers from severe limitations, hampering its use in practical uses. For example, phosphorene undergoes severe degradation phenomena in ordinary environments, leading to a very limited stability. Phosphorene is therefore very difficult to handle in real applications. In this paper, we briefly outline the concepts behind the application of predictive computational approaches to complex systems based on phosphorene, leading to the definition of practical strategies that can be applied to the development of stable, high-performance devices for electronics.

# Introduction

One of the most active fields of research in materials science concerns the development of materials for electronics. In this context, several novel classes of materials have been targeted, from molecular semiconducting materials, to conductive polymers and inorganic or hybrid low-dimensional systems. [1–6] Recent work has already demonstrated the potential of two-dimensional and layered compounds in the development of novel systems with peculiar structural, electronic, mechanical and magnetic properties to be exploited in practical applications. The most notable case is probably that of graphene, a two-dimensional allotrope of carbon.[7–13] However, other materials with a layered structure exhibit remarkable properties and a huge potential for applications.[4,14,15] Among these, black phosphorus, a layered allotrope of phosphorus, is one of the most promising. The interest in black phosphorus emerges especially in terms of its potential for electronic and optoelectronic applications.[16] In 2014, single-layer black phosphorus, or phosphorene, was also fabricated.[17]

However, a major drawback hampers the use of single- and few-layer black phosphorus and phosphorene in real life: as in many other allotropes, phosphorus layered structures react quite wildly with several agents in ordinary environments, such as water and oxygen.[18] As a result, black phosphorus and phosphorene are essentially useless in air or wet atmosphere, as they rapidly get oxidized. In other words, we have potentially great materials, but we cannot use them directly for applications in direct contact with air. We need therefore a strategy to protect phosphorene layers from external aggressive agents and, at the same time, keeping the outstanding properties of the active materials essentially unchanged. A possible strategy can rely on protecting the extremely sensitive phosphorene layers with inert, more atmosphere-resistant materials. Linear alkanes, for example, are essentially inert and interact weakly with other materials. The elongated structure of alkanes allows the formation of compact structures onto planar or pseudo-planar surfaces.[17,19] If coverage is efficient, the phosphorene and black phosphorus layers would be protected by a layer of alkanes, keeping most of its intrinsic properties essentially unchanged. This actually happens because long-chain alkanes tend to form well-aligned structures on several surfaces, as shown in Fig. 1. The process for protecting phosphorene and black phosphorus with alkanes could therefore be, at least in principle, quite easy: we just need to spread a thin layer of selected alkanes, which are in the liquid phase at room temperature, over the phosphorene upper layer. However, the development of practical strategies for enabling the use of phosphorene in applications requires the validation of these assumptions. In the following, we show how computational multiscale approaches can enable the development of predictive models of complex systems based on phosphorene, assisting experimental work in the realization of devices for applications.

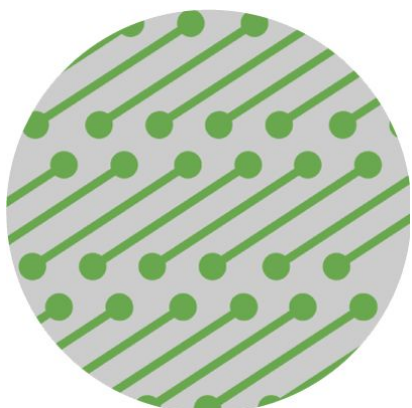

**Figure 1.** A compact arrangement of elongated objects (green) onto a surface (gray).

## Predictive simulations suggest how to protect phosphorene for applications

As mentioned before, multiscale computational modelling can deliver predictive information about the possibility of passivating phosphorene layers with alkanes, providing details about structure, electronic properties and effect of passivation in real applications. Atomistic molecular dynamics simulations show the effect of spreading liquid alkanes onto phosphorene and black phosphorus with liquid alkanes. We already applied molecular dynamics simulations with success in similar contexts for the study of materials for electronics.[20,21] The simulation protocols, methods and models need to be carefully tailored to address the specific conditions realized in the experiment.[22] One of the most appealing features of the simulation approaches proposed consist in the possibility of predicting the structure and properties of complex systems, materials and interfaces as a function of fabrication and growth conditions. This approach can therefore link basic materials properties with the experiments. In the case of phosphorene passivated by alkane layers, large-scale molecular dynamics simulations can correlate the morphology of alkane aggregates on the phosphorene surface with growth conditions, thus providing a "digital model" of what could happen in a real scenario.

The result is not trivial: in the experimental and simulation conditions, the long-chain alkanes considered are in the liquid phase, with an extremely disordered structure. The formation of large amorphous and unstructured aggregates of alkanes onto the phosphorene layer can therefore be expected. To our surprise, however, spreading linear alkanes (from C15 to C45 in this case) over phosphorene and black phosphorus resulted in the spontaneous formation of a well-packed protective layer. In Fig. 2, a representative snapshot extracted from simulations of a layer of alkanes in contact with the surface of phosphorene is shown. The packing of long chain alkanes on phosphorene is effective in a wide range of simulated growth conditions, thus suggesting the robustness of the passivation approach proposed. For alkanes on multi-layered phosphorene or black phosphorus the situation is quite similar.

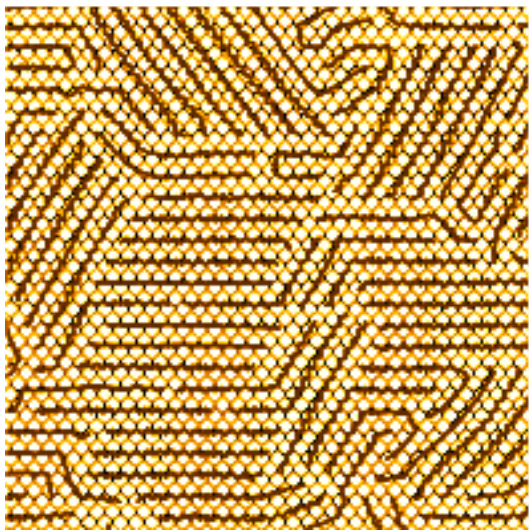

**Figure 2.** Snapshot of the molecular dynamics simulation of the formation of a layer of alkanes on phosphorene. Adapted from Ref. [23] with permission from the RSC.

As shown in Fig. 2, the alkane layer covers a large fraction of the exposed surface area. Moreover, DFT calculations confirmed that the interaction between alkanes and phosphorene or black phosphorus is quite weak. Surface passivation with alkanes should therefore affect only marginally the electronic properties of phosphorene and black phosphorus. These simulations therefore confirm the two most important points for using alkanes as protective layers: i) long chain alkanes form a compact layers on phosphorene, even at relatively low coverages, and are expected to prevent the surface from degradation; ii) passivation with alkanes keeps most of the attractive electronic properties of phosphorene essentially unchanged.

One of the most interesting applications of layered phosphorene compounds is as active layers in transistors. In ordinary transistor architectures, the active material (phosphorene, in this case) is supported by a dielectric, which keeps it apart from an electrode. It is therefore important to check what happens to phosphorene-based materials when in contact with a dielectric. Molecular dynamics simulations can predict the properties of passivated phosphorene supported on PMMA, a material often used as a dielectric layer in thin-film transistors.[21,24] These simulations tell us how to use phosphorene in a quite realistic scenario, and how the properties of phosphorene are possibly modified in this kind of environment. A representative snapshot of the simulation of a phosphorene layer sandwiched between a PMMA layer and alkanes is shown in Fig. 3. Simulations confirmed that alkane passivation is a very efficient route for preserving the properties of phosphorene also in the case of supported layers, pointing to the possibility of using phosphorene in practical applications.[23]

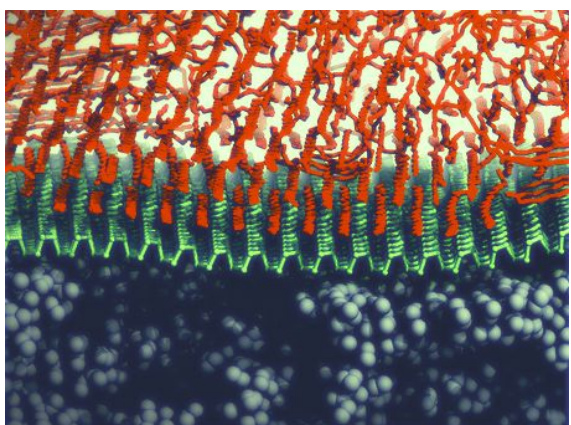

**Figure 3.** Snapshot of the molecular dynamics simulation of a phosphorene layer passivated by alkanes (red) and supported on PMMA (grey). Adapted from Ref. [23] with permission from the RSC.

Motivated by the computational predictions, experiments were also performed, confirming that long-chain alkanes form a compact layer on the surface, protecting phosphorene-based materials from oxidation: the regions where oxidation of phosphorene layers seem to occur are those that are apparently covered less efficiently with alkanes, as shown in Fig. 4. [25]

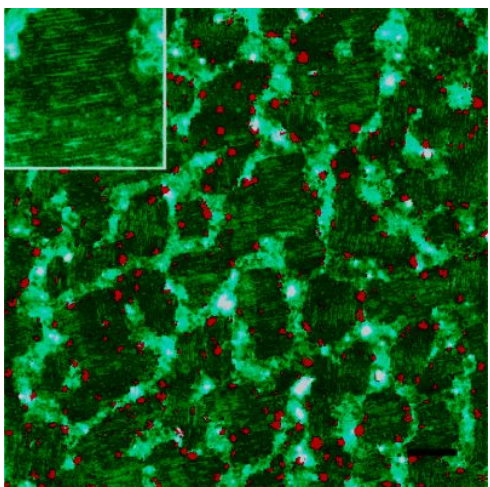

**Figure 4.** Atomic force microscopy image of a tetracosane layer on black phosphorus. The black scale bar is 100 nm. The red spot highlights zones where oxidation starts to occur after exposure to air. Adapted from Ref. [25] with permission from the RSC.

The experimental characterizations confirmed computational predictions, suggesting a practical strategy for the use of phosphorene-based materials in applications. Beside the technological implications of the application targeted, this work highlights the potential of predictive, multiscale computational approach in the development of advanced devices based on low-dimensional materials.

**Conclusions**

Despite the issues related to environmental stability, applications of two-dimensional phosphorus start to see the light. However, the properties of novel materials need to be tested in real-life scenarios. In this context, the full predictive power of computational modelling is unravelled, by providing properties of materials beyond ideality and models of suitable experimental processing. The predictivity of computational approaches relies on the definition of realistic model systems, which must take into account the complexity of the phenomena involved in the fabrication and working mechanism of devices. Large-scale simulations and interconnected chemico-physical models of complex systems can enable an unified picture of devices based on nanoscale materials across different scales, thus accelerating the development of novel devices for technological applications.